\documentclass[preprint,a4paper,11pt,review,numbers]{article}

\usepackage{graphicx}
\usepackage{amssymb}
\usepackage[english]{babel}
\usepackage[latin1]{inputenc}
\usepackage[numbers]{natbib}
\usepackage{amsmath}
\usepackage{amsfonts}
\usepackage{lineno}
\usepackage{authblk}
\usepackage[dvips]{pstcol}

\begin{document}

%
\title{The logical clarinet: numerical optimization of the geometry of woodwind instruments.}

\author[1]{D. Noreland}
\author[2]{J. Kergomard\thanks{Corresponding author: kergomard@lma.cnrs-mrs.fr}}
\author[3]{F. Lalo\"e}
\author[2]{C. Vergez}
\author[2]{P. Guillemain}
\author[2]{A. Guilloteau}

\affil[1]{Department of Computing Science, Ume\aa{} University,
Campustorget 5, 901 87 Ume\aa{}, Sweden}
\affil[2]{LMA, CNRS, UPR 7051, Aix-Marseille Univ, Centrale Marseille, F-13402 Marseille Cedex 20, France}
\affil[3]{LKB, ENS, CNRS, UPMC; 24 rue Lhomond, 75005 Paris, France}

\maketitle

\begin{abstract}
The tone hole geometry of a clarinet is optimized numerically. The instrument is modeled as a
network of one dimensional transmission line elements. For each (non-fork)
fingering, we first calculate the resonance frequencies of the input impedance peaks,
and compare them with the frequencies of a mathematically even chromatic scale
(equal temperament). A least square algorithm is then used to minimize the
differences and to derive the geometry of the instrument. Various
situations are studied, with and without dedicated register hole and/or enlargement of the bore. With a dedicated register hole, the differences can remain less than 10
musical cents throughout the whole usual range of a clarinet.
The positions,  diameters and lengths of the chimneys vary regularly over the
whole length of the instrument, in contrast with usual clarinets.
Nevertheless, we recover one usual feature of instruments, namely that
gradually larger tone holes occur when the distance to the reed increases.
A fully chromatic prototype instrument has been built to check these calculations,
and tested experimentally with an artificial
blowing machine, providing good agreement with the numerical predictions.
\end{abstract}


\section{Introduction}

Woodwind instruments of the orchestra have often attained their geometrical shapes
through a slow gradual process, which in many cases has taken
centuries.  Guided by trial and error, skilled craftsmen have managed to develop the instruments as we know
them today. In this article we study the clarinet.
Most of its evolutionary process (addition of new holes and keys, etc.)\ was made of the
succession of many small steps, each implying a limited departure from
a previous configuration --  for clarinets the only radical change was the introduction of the \textquotedblleft
Boehm system\textquotedblright\ of French instruments by Klos\'{e} in the middle of the 19$^{th}$ century. A typical wind instrument has
a large number of design parameters (positions and size of the holes
and the chimneys, bore, etc.), while many of them contribute at the same
time to the production of each note. Indeed, changing one of
them in order to correct a certain note may have an unexpected, and
often adverse, effect on other notes in terms of pitch, tone quality,
stability, etc. In a posthumous paper, Benade \cite{Benade-1994}
attempted to analyze the evolutionary path since the 18th
century.

Trying new configurations by the traditional method requires
a large amount of work. It therefore seems likely that the modifications
tested by the instrument makers have been limited to relatively
small changes, affecting only a few parameters at the same time. In
other words, in terms of optimization, existing
instrument designs probably represent local extrema of some optimization
function, in the sense that a small
change in the set of tone hole positions, radii etc. inevitably
worsens the instrument.  Nevertheless there might exist better geometrical shapes
that are more distant in the parameter space, and therefore not
accessible through small improvements of an existing design. An
additional reason to believe in this scenario is given by
the observation of the rather irregular tone hole pattern of many
woodwinds, with alternating small and large holes, short and long
chimneys, closed holes (opened for
one note only) etc. It seems that no particular physical principle could
explain why such an irregularity is desirable; there are actually reasons
to believe that it is not, in particular if homogeneity of the production of sound over the different notes is required.

Nowadays, with mathematical models of the instrument and computer
optimization algorithms, it is possible to test a number of configurations
that would be inaccessible by the traditional method. It is therefore
interesting to explore which results can be obtained by automatic optimization, to compare them with existing instruments, and to investigate if a
strong irregularity spontaneously emerges from the optimization. The idea
is not necessarily to create some completely new or exotic instrument, even
if this possibility is not excluded in the long run. It is rather to
investigate whether allowing large \textquotedblleft
leaps\textquotedblright\ from usual designs leads to a completely different
geometry of the instruments, to try and reach more \textquotedblleft
logical\textquotedblright\ configuration of the acoustical resonator, and
eventually test them acoustically. In particular, an open question (not answered in this work) is whether or
not the use of fork fingering, often used in clarinets, is an acoustical
necessity, or just the result of the complicated past
history of the instrument.

The purpose of this work is therefore to develop algorithms for
designing, and possibly improving, woodwind instruments, in the case of the
clarinet. It is to see if it is possible to conceive a
\textquotedblleft logical clarinet\textquotedblright,
with a perfectly regular fingering chart, and where the relations between the
acoustical functions of the resonator and its geometry are more easy to
grasp than in the traditional instrument. Of course, the instrument should
produce correct pitch for all notes. Fortunately this problem is not too
complicated to address in terms of calculated acoustical impedances: for simplicity it can be assumed that playing frequencies can be derived from
resonance frequencies with a simple length correction in order to account for reed flow and dynamics \cite{Nederveen}. A more difficult
issue is to design an instrument with balanced timbre over its entire
range. While the precise relation between tone quality and cutoff frequency
of the tone hole lattice \cite{moke:11} is still not
perfectly understood, experience seems to show that a regular cutoff frequency is useful (see \cite{ben:90}, page 485). Here, we study the possibility of
designing an instrument with a much more regular tone hole lattice in terms
of tone hole diameters and positions, able to produce a complete chromatic
scale over the full range of the traditional instrument.

Of course, whether such instruments will prove to be musically useful is not obvious a priori. Nevertheless, if this is
the case, it is clear that interesting perspectives for making simpler and
cheaper instruments could be envisaged. Our study is limited to the purely
acoustical aspects of instrument design; we have not studied the problem of
mechanical keys that are necessary for an instrumentalist to really play the instrument.
This is indeed an important question, but this task is beyond the scope of
the present work.

Numerous authors have discussed possible improvements of clarinets, in particular Benade \cite{Benade-1996}, but without using
numerical optimization. Brass instruments have indeed been studied by optimization
\cite{noudbe:10, kausel:0501, bradnewcamp:09}, but in this case the free parameters relate to the bore of the instrument and not to the geometry of lateral holes.

This article is organized as follows. Section \ref{sec:mathmodel} provides
the basic mathematical model used to characterize the acoustical properties
of the instrument -- mostly a calculation of the resonance frequencies of
the resonator. Section \ref{sec:opt_proc} describes the optimization
procedure and the minimization algorithm. Section \ref{sec:compimpl}
briefly discusses the computer implementation. Section \ref{sec:prospective}
presents various numerical results obtained by retaining various
optimization criteria; five different \textquotedblleft
clarinets\textquotedblright\ are obtained and their properties are
compared. These results are used in section \ref{sec:prototype} to design an
experimental prototype, and to measure its sound production with the help of
an automatic blowing machine. Finally,
section \ref{sec:discussion} draws a few conclusions.

\section{Mathematical model \label{sec:mathmodel}}

\subsection{Transmission line model}

The instrument is modeled with a classical one-dimensional transmission
line model for planar waves \cite{keefe:90}, taking visco-thermal losses
into account throughout the main bore, as well as in the tone holes. It is
assumed that the distance between tone holes is sufficiently large to make higher mode
interactions negligible.
This assumption is valid if
the distance is at least larger than the bore diameter (see e.g. Ref. \cite{kegatada:89}).
Accordingly, the instrument is modeled as a succession of transfer matrices
representing either a cylindrical piece of tubing, or a tone hole; each tone
hole is formally represented by a lumped element.

\begin{figure}[htb]
\begin{center}
\leavevmode
\setlength{\unitlength}{1mm}
 \includegraphics[width=7cm]{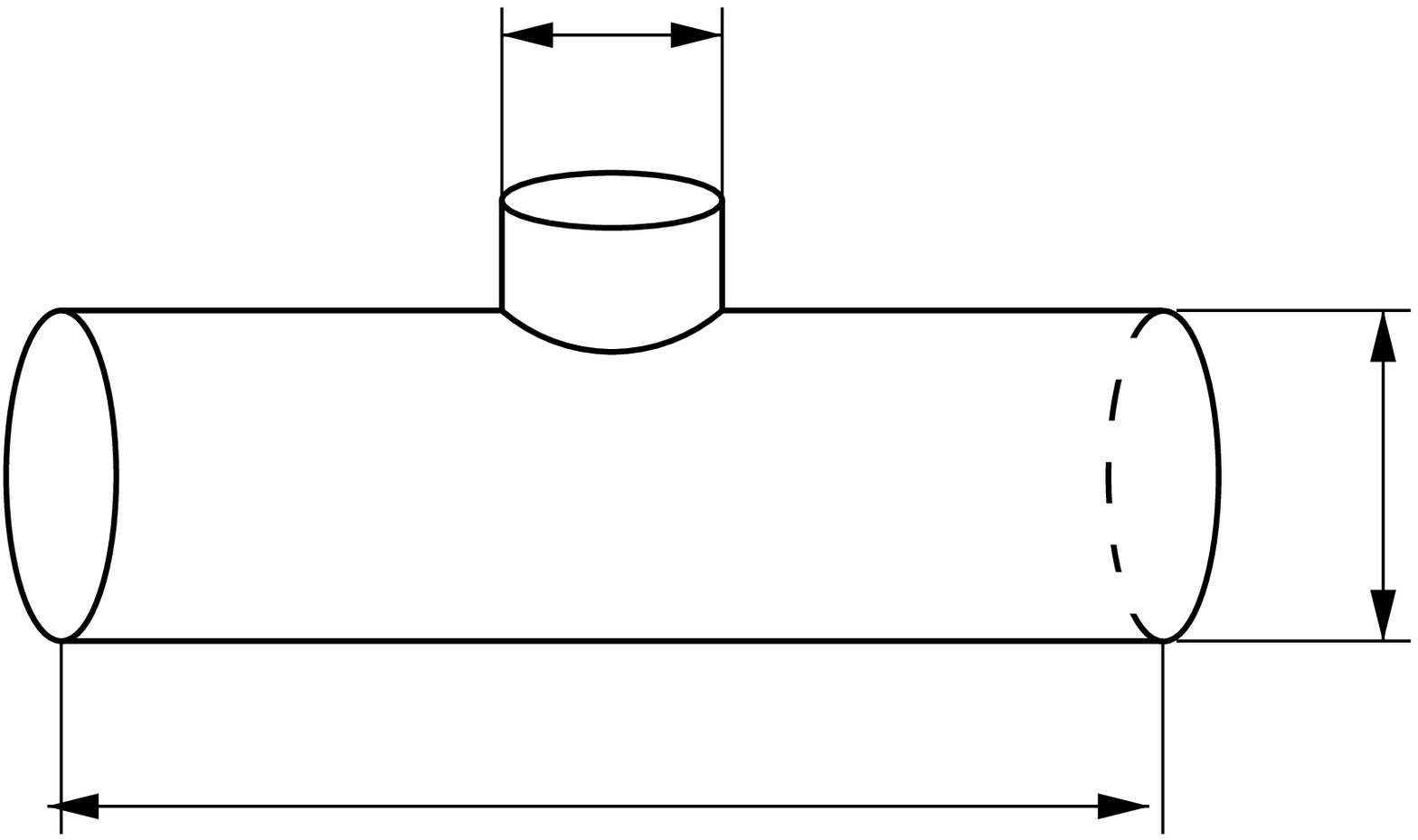}
\put(-42.5,44){$2b$} \put(0,17){$2a$} \put(-42.5,5){$2d$}
\end{center}
\caption{Elementary cell with tone hole.}
\label{fig:cell}
\end{figure}

The transfer matrix of a cylindrical piece of tubing of length $L$ and
characteristic impedance $Z_{c}$ is given by
\begin{equation}
H=\left[
\begin{array}{cc}
\cosh (\Gamma L) & Z_{c}\sinh (\Gamma L) \\
(1/Z_{c})\sinh (\Gamma L) & \cosh (\Gamma L)%
\end{array}%
\right],
\end{equation}%
where $\Gamma $ is the complex propagation constant. The model is rather
accurate for the characteristic wavelengths propagating inside a typical wind
instrument. The first higher order mode is usually far below cutoff; for a
cylinder of 15mm diameter it is a helical mode with a cutoff frequency of 13.5kHz.

\subsection{Visco-thermal boundary layer effects}
\label{visco-thermal}

The following expressions for the characteristic impedance $Z_{c}$ and the
wave number $\Gamma $ are used (see e.g. \cite{keefe:84}) 
\begin{align}
Z_{c}& =Z_{0}\left[ \left( 1+\frac{0.369}{r_{v}}\right) -j\left( \frac{0.369%
}{r_{v}}+\frac{1.149}{r_{v}^{2}}\right) \right], \\
\Gamma & =k\left( \frac{1.045}{r_{v}}+\frac{1.080}{r_{v}^{2}}+j\left( 1+%
\frac{1.045}{r_{v}}\right) \right).
\end{align}
In this equation, $Z_0$ is equal to
\begin{equation}
Z_{0}=\frac {\rho c}{\pi a^{2}},
\label{Z0}
\end{equation}
where $\rho$ is the mass density of the gas, $c$ the speed of sound
and $a$ the radius of the tube. $k$ denotes the
wavenumber $\omega/c$, where $\omega$ is the angular frequency. The
dimensionless number $r_{v}$ is defined as the ratio between the tube
radius and the thickness of the boundary layer
\begin{equation}
r_{v}=a\sqrt{\rho \omega /\eta },
\end{equation}%
where $\eta $ is the coefficient of viscosity.

\subsection{Tone holes}

\begin{figure}[htb]
\begin{center}
\leavevmode
\setlength{\unitlength}{1mm} \includegraphics[width=6cm]{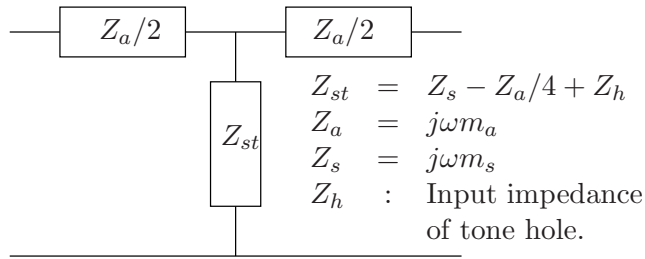}
\put(-48,29){$Z_a/2$} \put(-20,29){$Z_a/2$} \put(-32,15){$Z_{st}$}
\put(-22,12){$\begin{array}{lcl} Z_{st}&=&Z_s-Z_a/4+Z_h\\Z_a&=&j\omega m_a\\
Z_s&=&j\omega m_s\\Z_h&:&\mbox{Input impedance}\\ &&\mbox{of tone hole.}
\end{array} $}
\end{center}
\caption{T-circuit equivalent for a tone hole.}
\label{fig:Tjunction}
\end{figure}

Each tone hole is modeled as a T-junction (Fig. \ref{fig:Tjunction}).

The transfer matrix corresponding to this electrical equivalent circuit is the
following, if $Y_{s}=1/(Z_{s}+Z_{h})$:%
\begin{equation}
\frac{1}{1-Y_{s}Z_{a}/4}\left(
\begin{array}{cc}
1+Y_{s}Z_{a}/4 & Z_{a} \\
Y_{s} & 1+Y_{s}Z_{a}/4%
\end{array}%
\right).
\end{equation}

The series impedances $Z_a/2$ are purely inertial, but the
total shunt impedance $Z_{st}$ also has a resistive part due to
visco-thermal damping and radiation losses. For the acoustic masses $m_a$
and $m_s$, we use expressions obtained from \cite{dubos:99,dalmont:02},
\begin{align}
m_a &= \rho t_a/(\pi a^2), \\
m_s &= \rho t_s/(\pi b^2),
\end{align}
where
\begin{align}
t_s &= b(0.82-0.193\delta-1.09\delta^2+1.27\delta^3-0.71\delta^4), \\
t_a &= b(-0.37+0.087\delta)\delta^2, \\
\delta &= b/a.
\end{align}

The input impedance $Z_{h}$ of a tone hole of cross section area
$S_{h}=\pi b^{2}$ depends on whether it is open or closed. For an open
tone hole, $Z_{h}$ is calculated by considering the tone hole as a
transmission line terminated by a radiation impedance $z_{L}$. A
simple expression for the radiation impedance of a hole in the side of
a cylinder \cite{dalnedjol:01} is not known but, since $ka$ is small,
it seems reasonable to assume that the tone hole acts as an infinitely
flanged pipe; a more detailed model for flanged termination is
probably unnecessary for our purposes. At low frequencies
($ka \ll 1$), this leads to the simple formula%
\begin{equation}
z_{L}=\frac{\rho c}{S_{h}}\left[ \frac{1}{4}(ka)^{2}+j0.82ka\right].
\end{equation}
Accordingly, a tone hole of length $h$, terminated by an impedance $z_{L}$,
is represented by the input impedance
\begin{equation}
Z_{h}=\frac{\rho c}{S_{h}}\frac{z_{L}+j\frac{\rho c}{S_{h}}\tan (kh)}{\frac{%
\rho c}{S_{h}}+jz_{L}\tan (kh)}.
\end{equation}%
Exterior hole interaction \cite{kergomard:89} is not taken into
account; assuming that this effect remains negligible is reasonable,
especially at low frequencies. The input impedance of a closed tone
hole is calculated in the same way, but with $z_{L}\rightarrow \infty
$. In the limit $kh\ll 1$, which is an acceptable approximation of the
impedance for short chimneys, the closed hole input
impedance expression reduces to a shunt stiffness $\rho
c^{2}/(j\omega S_{h}h)$.

\subsection{Termination of the instrument} \label{sec:termination}

\begin{figure}[tbh]
\begin{center}
\leavevmode
\setlength{\unitlength}{1mm} \includegraphics[width=6cm]{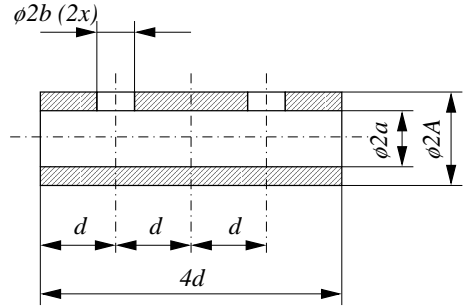}
\end{center}
\caption{Instead of a bell, the main tubing is extended and fitted with two
vent-holes. For $2a=$14.75 mm, $A=$12.5 mm, $b=$4.0 mm, and $d=$18.2 mm the
cutoff frequency $f_{c}=$1.420 kHz.}
\label{fig:ventholes1500}
\end{figure}

An ordinary clarinet is terminated by a bell. The main purpose of the
bell is to equilibrate the timbre of the lowest notes of the
instrument with that of the other notes. In this project, we replace
the bell by a continuation of the cylindrical main bore with two
vent-holes, as shown in Fig. \ref{fig:ventholes1500}.  The length of
the extension and the diameters of the vent-holes are chosen in order
to obtain a theoretical lattice cutoff frequency of 1.420 kHz,
approximately equal to the average cutoff frequency of a clarinet
\cite{moke:11}.

\subsection{Calculation of playing frequencies}

The frequency of a blown note depends on the input impedance spectrum,
the reed dynamics (in contact with the lips) and the blowing
pressure. In practice, since the playing frequency is much smaller
than the resonance frequency of the reed, the dominant factor is the
input impedance.

We use a simplified description where the mouthpiece and the reed are
replaced (for a given stiffness, blowing pressure, air flow
correction, etc.) by an effective volume correction added to the
instrument.

 The effects of a temperature gradient along the air
column on the pitch can be ignored in a first approximation. References
\cite{WaCuCa:97} and \cite{GiLeGo:06} indicate that a satisfactory
approximation of the frequency shifts is obtained by taking into
account the average along the instrument of the temperature only; our
calculations should then remain valid, just with a temperature
offset. Another approximation we have made is not taking into account
the effects of changes in gas composition (CO$_2$, O$_2$), which have
been studied by Coltman\cite{coltman:79} for the flute and
Fuks\cite{fuks:97} for the oboe and bassoon. Further investigation is
probably needed along these lines. 

The playing frequency at soft playing levels is to a first
approximation equal to a resonance frequency of the air column inside
the clarinet-mouthpiece combination, given by the solution of the
equation
\begin{equation}
\text{Im} [Z_{in}(\omega )]=0\text{, }  \label{eq:imzinzero}
\end{equation}%
where $Z_{in}$ is the input impedance of the whole instrument at the
tip of the reed, including the volume correction of the mouthpiece and
the reed.

Apart from the resonance frequencies of the resonator, of course other
properties of the input impedance spectrum may influence sound
production. In particular, the heights and widths of the impedance
peaks are relevant to the stability of played notes. The importance of
aligning the harmonics of the playing frequency with subsequent zeros
of Eq.~\ref{eq:imzinzero} has been emphasized by Benade \cite{ben:90}.
According to a theoretical evaluation (\cite{chake:08}), an
inharmonicity of 20 cents between the two first peaks may cause a
variation in the playing frequency of about 10 cents between piano and
fortissimo levels. Moreover, transients during the attack of the notes
may be affected by other properties of the impedance
spectrum. Nevertheless, since less deviation of intonation can be
tolerated during the quasi permanent regime of sounds, we have chosen
to include only the resonance frequencies of the input impedance in
our optimization.

\section{Optimization procedure\label{sec:opt_proc}}

\subsection{Cost function and minimization algorithm}

The principle of clarinet design optimization is to determine a set of
geometrical variables that minimize a cost function characterizing,
for each fingering, the distance between the solutions of
Eq. (\ref{eq:imzinzero}) and the frequencies of a tempered
scale. Since the number of design variables is large, and since the
cost function depends non-linearly on them, a numerical treatment of
the problem with an efficient minimization algorithm is necessary. We
have chosen gradient based algorithms for their convenience; they do
not guarantee to reach the absolute extremum in general, but are
efficient to find local optima.

To start the algorithm, a reasonable initial guess for tone hole
positions and dimensions is necessary, as a \textquotedblleft
seed\textquotedblright\ for the calculation.  This seed was obtained
by starting from the lowest note, which gives the total length of the
instrument, and then successively computing by iteration the position
for each tone hole in order to obtain the desired resonance
frequencies of the first register.  If the radius and
chimney length of each hole are fixed to some typical value
(in the examples below, the radius increases linearly
from 4.0 mm to 5.0 mm and the chimney length is 4.0 mm), and if the influence
of closed tone holes above the first open one is ignored, the process
amounts to solving a series of scalar equations for the hole
positions.

The cost function was calculated by taking into account the
frequencies $f_q^k$ of the impedance resonances obtained from
Eq. (\ref{eq:imzinzero}), where $k$ refers to the note ($k=1,2,..,
N_{notes}$) and $q$ refers to the resonance ($q=1$ corresponds to the
first impedance resonance, $q=2$ to the second, etc.). For the lower
register, the cost function includes two elements: the square of the
distance between the first impedance resonance and the frequency
$\tilde{f}^k$ of a tempered scale, as well as the square of the
distance between the second resonance and $3\tilde{f}^k$, both with
equal weights. In this way, a good impedance peak cooperation can be
expected, resulting in good pitch stability.  For the second register,
only the first resonances $f_2^k$ were taken into account and compared
to the corresponding equal scale values $\tilde{f}^k$.  In practice,
we introduce a vector $\mathbf{R}$ with $2N_1$ components associated
with the lower register (where $N_1=19$, the number of notes of this
register):
\begin{equation}
\begin{array}{lcl}
R_{2k-1} & = & (\tilde{f}^k-f_1^k) / \tilde{f}^k, \\
R_{2k} & = & (3\tilde{f}^k-f_2^k) / (3\tilde{f}^k),%
\end{array} \label{eq:residuals}
\end{equation}
as well as $N_2$ additional components associated with the second
register. We then choose the square of the norm of the vector
$\mathbf{R}$ with $2N_1+N_2$ components as our target
function for optimization: $F\equiv\mathbf{R}\cdot\mathbf{R}$.
With $\mathbf{x}$ representing a vector of physical design variables as described below, the optimization problem can be stated as
\[ \min_\mathbf{x} F(\mathbf{x}) \qquad
\textrm{subject to constraints on $\mathbf{x}$}.\]

The problem is expected to be non-convex, leading to many extrema that
 are in general only local, and therefore dependent on the seed of the
 calculation. Nevertheless, the hope is that the crude initial model
 of the clarinet used to create this seed should be sufficiently
 reasonable to make a sensible instrument emerge from the
 optimization.

It is probably impossible to attain $F=0$ (simultaneous perfect
position of resonances for all considered notes). What is obtained is
a compromise, which can then be adjusted if necessary by weighting the
terms of the cost function differently. For instance, the even
components of $\mathbf{R}$ corresponding to the second resonance of
the first register may be considered as less important than the odd
components.

\subsection{Design variables}

The free parameters $\mathbf{x}$ of the model are the
total tube length and the positions, radii and chimney lengths of the
tone holes, which amounts to more than $50$ free parameters. The
resonator is perfectly cylindrical; nevertheless, a localized
cylindrical enlargement/constriction between the mouthpiece and the
uppermost tone hole can also be introduced into the calculation, since
this is known to improve harmonicity \cite{dekela:05}.  We also put
constraints on the tone hole diameters and chimney lengths, in order
to avoid unpractical or otherwise unfeasible
solutions. Some constraints are straightforward (such as dimensions
being positive, and the hole radii necessarily being smaller than the
radius of the main bore), but others are required by manufacturing, or
by the fact that the mathematical model would otherwise not be valid.
In practice, those constraints were often left for manual a posteriori
check.

All variables do not affect the distances of Eqs. (\ref{eq:residuals})
in the same way. For instance, it is obvious that the holes of the
bottom notes have little influence on the tuning of the upper
resonances. On the other hand, the uppermost tone holes generally have
an appreciable influence on all of the lower notes, due to the shunt
reactance introduced by closed tone holes.

As a simple first approximation, the effect of an open
tone hole of length $h$ can be represented by a shunt acoustic mass
$M_h = \rho(h+1.6b)/\pi b^2 $, which suggests that $h$ and $b$ do not need
to be simultaneously considered as design variables. In practice, however, it
appears necessary to include also the chimney lengths as design variables
in order to obtain acceptable positions of the resonances.

\section{Computer implementation \label{sec:compimpl}}

The core of the algorithm is the calculation of a function giving the
input impedance of a series of open and closed tone holes, separated
by cylindrical sections. It is used by a routine that evaluates
$\mathbf{R}$ and the cost function, using a global root finder in the
search for the zeros of $\text{Im} (Z_{in})$.  The global root search
is essentially done by analyzing the spectrum and selecting out the
impedance maxima of interest before Eq.~\eqref{eq:imzinzero} is
solved.

\begin{figure}[tbp]
\begin{center}
\leavevmode
\includegraphics[width=5cm]{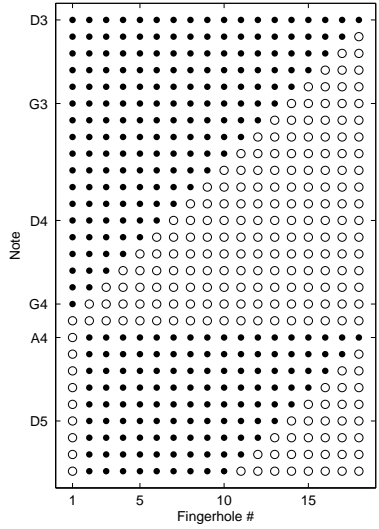}
\end{center}
\caption{Fingering chart for a chromatic instrument where the G\#4 hole also
serves as the register hole.}
\label{fig:fingerchart1}
\end{figure}

One of its input of the optimization code is a fingering-matrix (such
as that shown in Fig.~\ref{fig:fingerchart1}). This makes optimization
with arbitrary fingerings possible, for instance even if fork
fingering was considered. The algorithms are implemented in Matlab,
and the routine \verb\lsqnonlin\ from the optimization toolbox is used
for the optimization procedure, with the necessary
gradient approximated numerically. The stopping criterion for the
iteration (change in $F<10^{-8}$) was chosen empirically in accordance
with the magnitude of $F$ (see the convergence study in
\ref{sec:convstudy}). The execution time for the optimization depends
on the number of design variables, the number of components in the
cost function and the convergence process. For the designs presented
below, it varied typically between 20 min and two hours on a desktop
computer.

\section{Various designs} \label{sec:prospective}

A first series of numerical experiments was made in order to get a
better idea of a suitable configuration in general; five different
configurations, denoted \emph{a}--\emph{e}, were investigated. They
all represent strictly chromatic instruments, meaning that they
include neither fork fingerings nor ``disordered'' opening of tone
holes.  The frequency range was D3--F5, corresponding to the first
register (chalumeau) and the first $N_2=10$ notes of
the second register -- for case (e), it was even slightly more, as we
discuss below.  The clarinets differ in the function of the register
hole, which can be either a dedicated register hole, or a dual
register hole/tone hole. In addition, the effect of a cylindrical
constriction or enlargement between the mouthpiece and the first tone
hole was investigated -- this was the only deviation from an otherwise
cylindrical bore.

A cylindrical instrument such as the clarinet overblows the
twelfth. For a chromatic instrument, this requires 18 tone holes to
cover the range of the first register. The notes of the second
register
are obtained by opening the register hole and repeating the fingering
from the first register. A fingering chart for the instruments with a
dual register hole can be seen in Fig.~\ref{fig:fingerchart1}.

Initially, the bore diameter $2a=14.75$ mm was selected to match
available clarinet mouthpieces; $a$ was therefore not considered as a
variable in the optimization. The dimension $d$ of the instrument
termination (Fig. \ref{fig:ventholes1500}) was then calculated from
$a$ and the chosen values $b=4.0$ mm and $A-a=5.0$ mm so as to give a
cutoff frequency of 1.42 kHz (on real clarinets, the chimney length $h$ is often longer than the tube thickness $A-a=$). The constraints imposed on the hole
dimensions were rather loose with respect to the values of existing
instruments. A lower bound on the hole radius was set to 2.0 mm,
except for the tone hole acting as the register hole, for which it was
set to 1.0 mm. An upper bound of 6.0 mm was set for all holes.

For chimney lengths, a lower bound was set to 2.5 mm, with no upper
bound.  If one includes a dedicated register hole, 20 cylindrical
sections precede, separate, or succeed the 19 tone holes. Each tone
hole is characterized by two parameters, which now makes a total of 58
design variables.

To achieve convergence, it proved necessary to perform
the optimization process in two successive phases. Starting from the
crude initial solution described in section \ref{sec:opt_proc}, the
optimization process was run by calculating a single-register design
($N_2=0$) optimizing only the 19 notes of the first register
(including their second resonances). This solution is then used as the
starting point for phase 2, which takes into account the second
register also.  Experience shows that phase 2 is more sensitive to the
initial solution than phase 1. A ``bad'' initial solution might in
practice ruin convergence altogether, or lead to a local minimum that
is clearly not acceptable.

\subsection{No specific register hole -- case (a)}
Our first optimization was the design of a clarinet with an uppermost
tone hole that has the dual function of a register hole and an
ordinary tone hole, as common with existing instruments.  The role of
the register hole is to shift and reduce the height of the peak of the
fundamental resonance, while the second resonance is not too affected;
this facilitates the emission of the second register. These conditions
tend to lead to register holes that are significantly smaller than
tone holes, so that some compromise is necessary for a hole having a
dual function.

\begin{figure}[tbp]
\begin{center}
\leavevmode
\includegraphics[width=11cm]{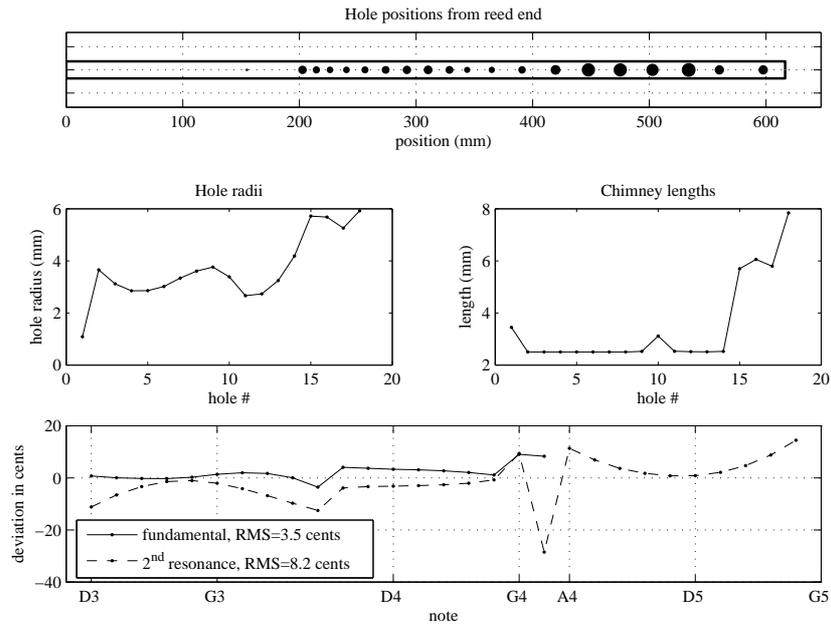}
\end{center}
\caption{Design (a), corresponding to a clarinet with no dedicated
register hole.  The upper part of the figure shows the positions of
the 18 tone holes; the two last holes on the right correspond to the
acoustical lattice replacing the bell, and have not been optimized
(see \S\ \ref{sec:termination}).  The two intermediate figures give
more detail on the geometry of these tone holes. The lowest part of
the figure shows the difference between the position of the impedance
resonances and the frequencies of a perfect chromatic
scale with equal temperament. Notes 1--19 are the first (lower)
register notes, notes 20--29 are second register notes calculated from
the second resonance of the impedance.  One notices the particular
position of the first hole, which is unusually separated from all the
others; this is a consequence of its dual acoustical role (register
and tone hole).}
\label{fig:CylNoreg}
\end{figure}

Fig.~\ref{fig:CylNoreg} shows the obtained tone hole pattern and the
position of the acoustical resonances with respect to equal
temperament. The position of the first tone hole/register hole is 155
mm from the reed end, which is about one third of the distance to the
F3 hole; this is near the optimal position of a register hole for the
bottom notes of the second register. The hole radius is 1.09 mm, which
is in fact slightly larger than the constraint, and considerably
smaller than all the other tone holes. This explains the large
distance between the first and the second holes compared to the rest
of the tone hole lattice. The first resonances of the notes in the
first register are well in tune. Around the crossover from the first
to the second register, resonance tuning problems occur.

\subsection{Adding a cylindrical enlargement to the bore -- case (b)}

Introducing a cylindrical enlargement is known to correct the tuning
of the twelfths \cite{dekela:05}. Our optimization code is compatible
with the introduction of a cylindrical constriction or enlargement
anywhere between the mouthpiece and the uppermost tone hole. As
mentioned above, the latter is represented by a section of the
resonator having the same volume as a typical mouthpiece (for a 14.75
mm bore this corresponds to 73 mm.). Optimizing this
constriction/enlargement introduces new parameters: its position,
length and diameter. An upper bound on the diameter was set to 25.0
mm.  Lower bounds for both the diameter and the length
were set to 0 mm.

The optimization provided a 4.4 mm long enlargement with a diameter of
25 mm, inserted immediately after the mouthpiece. The diameter was
therefore equal to its maximal bound, introducing a rather large
discontinuity; under these conditions, higher order duct modes should
be taken into account, introducing added mass \cite{kega:87}. As a
simple approximation, it can be considered as a simple length
correction, found to be 1.5 mm.

\begin{figure}[tbp]
\begin{center}
\leavevmode
\includegraphics[width=11cm]{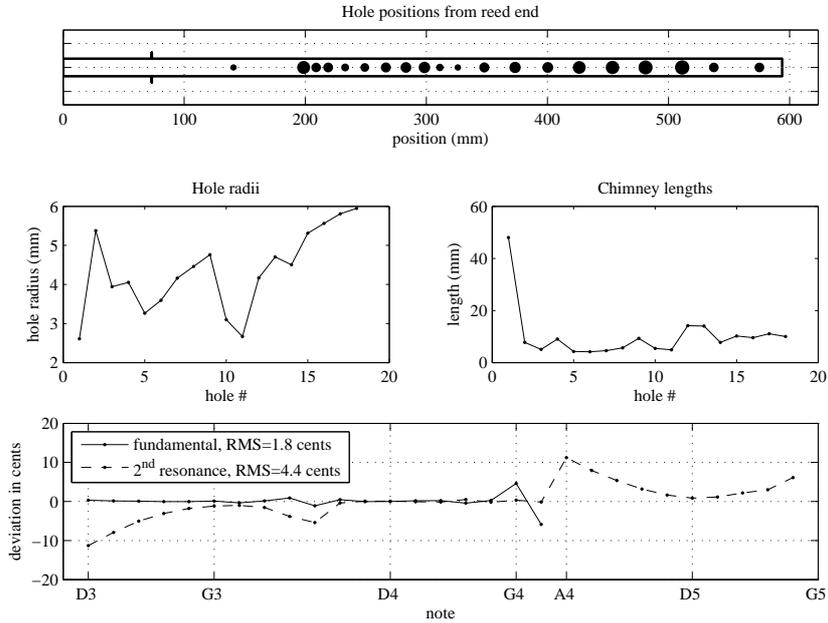}
\end{center}
\caption{Design (b), corresponding to a clarinet with a bore
enlargement but no dedicated register hole.  The different parts of
the figure are defined as explained in the caption of
Fig. \ref{fig:CylNoreg}.  The enlargement has a length of 4.4 mm and a
diameter of 25 mm, put immediately after the mouthpiece. One notices
one unusually long first hole (about 50 mm), which seems a rather
impractical value for a hole intended to emit
sound. }
\label{fig:EnlargNoreg}
\end{figure}

Fig.~\ref{fig:EnlargNoreg} shows the positions of the holes as well as
the obtained positions of the resonances with respect to equal
temperament. A comparison with Fig.~\ref{fig:CylNoreg} shows that the
addition of the bore enlargement has already introduced a significant
improvement.

\subsection{Specific register hole with cylindrical bore -- case (c)}

The use of a separate register hole removes one important acoustic
compromise concerning its size. But it is well known that a compromise
is still necessary concerning its position, since a register hole
should be ideally placed at a pressure node of each note, which is of
course impossible to obtain simultaneously for all of them. The role
of the optimization is precisely to find this compromise.  We note,
nevertheless, that it does not take into account the height of the
resonance peaks; the position of the register hole is only determined
by the positions of the second resonances (and, of course, by
constraints as well).

\begin{figure}[tbp]
\begin{center}
\leavevmode
\includegraphics[width=11cm]{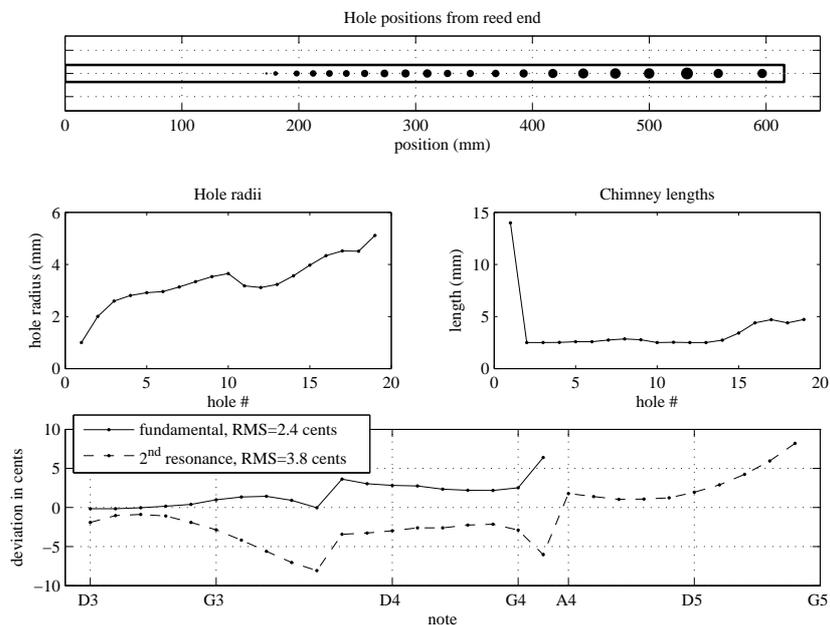}
\end{center}
\caption{Design (c) with a specific register hole,
but no bore enlargement. The different parts of the figure are defined
as explained in the caption of Fig. \ref{fig:CylNoreg}.
Here, the length of the
register hole, about 15 mm, is comparable to that of real clarinets.}
\label{fig:CylReg}
\end{figure}

Fig.~\ref{fig:CylReg} shows the results. Compared to configuration (a)
with a dual register hole, a more even tone hole progression is
achieved, while at the same time the frequency differences are
reduced. The position of the register hole is roughly at one third of
the position to the tone holes of the bottom notes, making it optimal
for the first notes of the second register. Its diameter reaches the
minimum radius 1.00 mm allowed by the constraint. The constraint
concerning hole no. 2 (the first tone hole) also determines its radius
of 2.00 mm; all the other holes have a size that remains between the
bounds. The constraint of 2.50 mm for the length of the chimneys is
active for some of the holes, but the variation for the rest of the
holes is rather smooth.

\subsection{Combining specific register
hole and cylindrical enlargement -- case (d)}

\begin{figure}[tbp]
\begin{center}
\leavevmode
\includegraphics[width=11cm]{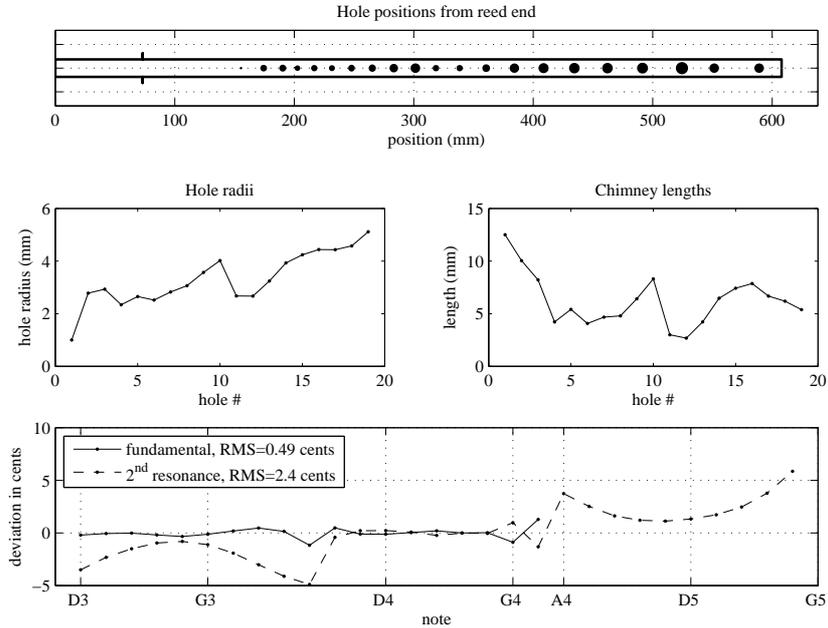}
\end{center}
\caption{Design (d) with both a specific register hole
and a bore enlargement just after the mouthpiece. The different parts of the figure
are defined as explained in the caption of Fig. \ref{fig:CylNoreg}.}
\label{fig:EnlargReg}
\end{figure}

Adding a bore enlargement to the design with a separate register hole
improves intonation further.  As in design (b), we put an upper limit
of $25$mm on the maximum bore diameter, and the enlargement is put
directly after the mouthpiece. Optimization reached this maximum and
provided a length of 1.9 mm.  Fig.~\ref{fig:EnlargReg} shows the
results. The fundamental register is now in tune within 0.5 cents RMS;
only the highest note is out of tune by more than 5 cents, which is
still a very small shift.

\subsection{Complete second register -- case (e)}

\begin{figure}[tbp]
\begin{center}

\leavevmode
\includegraphics[width=10cm]{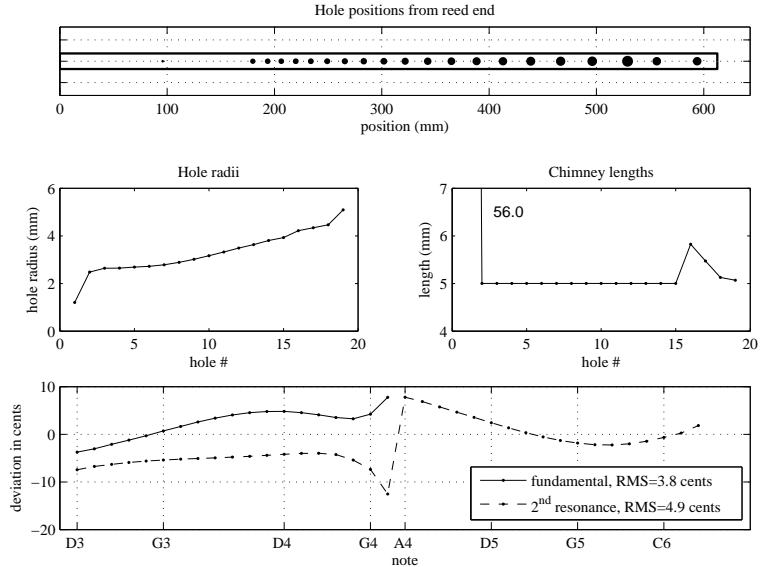}
\end{center}
\caption{ Design (e) of an instrument covering three octaves including the upper notes of the second register. The different parts of the figure are defined as explained in the caption of Fig \ref{fig:CylNoreg}. The register hole in this case turns out to be rather long, namely 56 mm.}
\label{fig:expclar1geom}
\end{figure}

\begin{figure}[tbp]
\begin{center}
\leavevmode
\includegraphics[width=10cm]{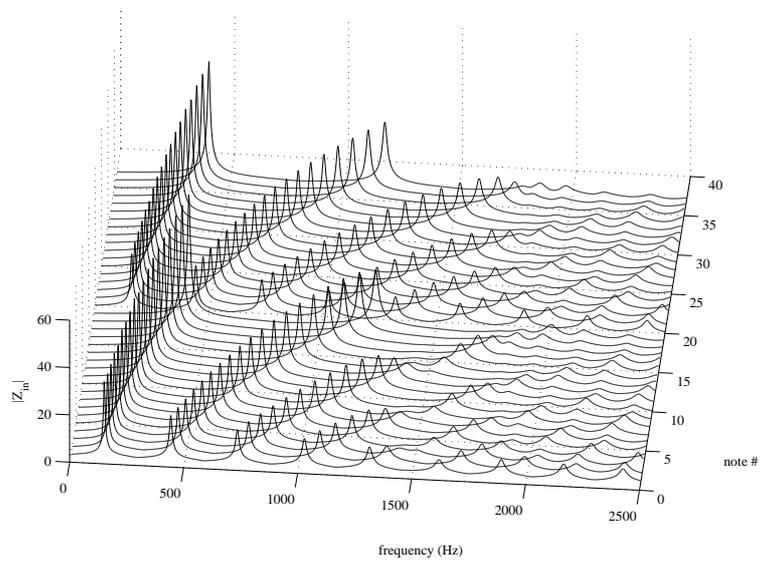}
\end{center}
\caption{Computed input impedance spectra for the 37 notes of design
(e). The impedance is made dimensionless by dividing it by the
characteristic impedance $Z_{0}$ defined in \S\ \ref{visco-thermal}.
The impedance peaks are similar to those of a real clarinet but, for
the second register, the second peak remains smaller than the first
one, because of the long register hole.}
\label{fig:expclar1WF}
\end{figure}

Finally, we studied a 5th case, clarinet (e). Among candidates
(a)-(d), clarinet (c) seems to provide the best compromise in terms of
intonation and geometrical regularity; we then decided to extend the
study of this design by exploring the possibility of tuning resonances
of all notes of the second register -- including the highest notes,
which are normally played with the third register of standard
instruments. In this case, two registers cover three full octaves
$D_3$--$D_6$, where the second resonance is used throughout the second
register ($N_2=18$). Since the frequencies of the
highest notes are approaching the resonance of the reed (around 2
kHz), it is likely that the assumption of blown notes having
frequencies equal to impedance resonances is less accurate in the
highest part of the second register \cite{sikevegi:08}. Nevertheless,
it is known that real clarinets provide a rather large pitch
flexibility in the high register; small errors in this range should
not be too problematic. The position of the register hole was subject
to a constraint of a maximal distance of 100 mm from the mouthpiece
end, chosen to render the hole effective also in the
upper part of the second register. For this design, a bore diameter
$2a=$ 14.27 mm was chosen (instead of 14.75 mm for the other designs)
to better correspond to the experiments described in the next section.
Fig.~\ref{fig:expclar1geom} shows the results for this design,
with a long register hole, and
Fig.~\ref{fig:expclar1WF} shows the computed impedance spectra
associated with it.

\begin{figure}[tbp]
\begin{center}
\leavevmode
\includegraphics[width=10cm]{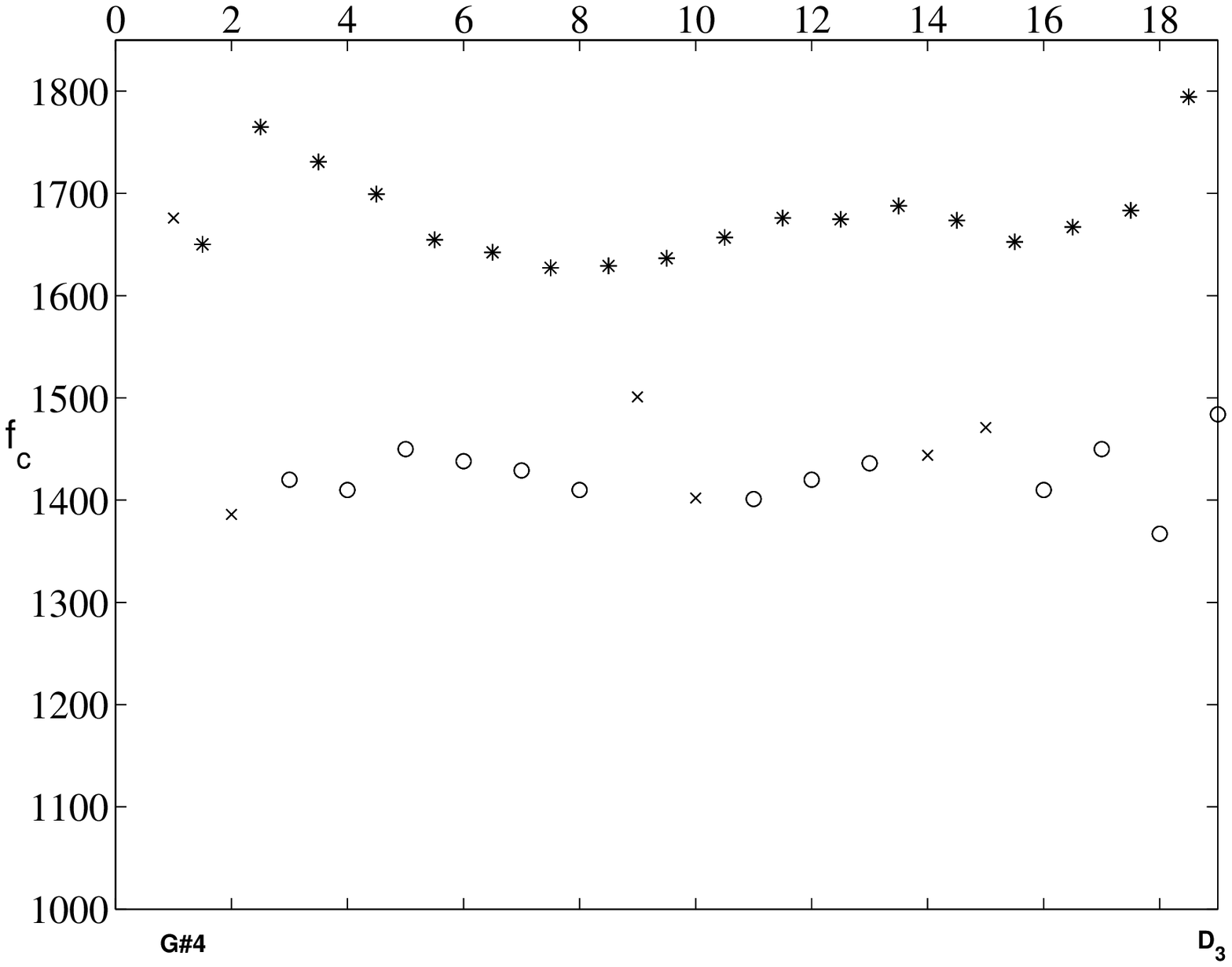}
\end{center}
\caption{The local cutoff frequencies for a set of two holes are shown
with stars (*). A star located at $n+1/2$ corresponds to the cutoff
frequency of the set of two holes ($n$, $n+1)$. The star located at
$18.5$ is calculated for the hole 18 and the first vent-hole.\newline
The circles and crosses represent the global cutoff frequencies
obtained in \S\ \ref{sec:prototype} from the measurement of the input
impedance, for the notes of the first register from D3 to G\#4.
Circles correspond to well defined
values, crosses to more uncertain values.}
\label{fig:cutoff}
\end{figure}

For this design, we have also studied the acoustical regularity of the
lattice of tone holes \cite{moke:11}. Local cutoff frequencies of
$\Pi$-shaped sections can be considered as a criterion of acoustical
regularity: if these frequencies remain constant over the various
holes, the instrument should behave as a periodic lattice with the
corresponding cutoff frequency, and should therefore provide a better
homogeneity of sound production.  The computed local cutoff
frequencies of the $\Pi$-shaped sections for 18 tone holes are shown
by the stars in Fig.~\ref{fig:cutoff}. The relative variations of the
cutoff frequency are about 10\%, while standard clarinets have a
variation of the order of 40\%.  Therefore the computed clarinet has a
satisfactory acoustical regularity of its acoustical lattice. As for a
real clarinet, the mean value of the local cutoff frequencies lies
around $1700$ Hz. This is significantly higher than the global cutoff
frequencies measured from the input impedance curve for the notes of
the first register, which is around $1450$ Hz as shown in
Fig.~\ref{fig:cutoff}. This discrepancy illustrates the difficulty of
defining and measuring global cutoff frequencies for a regular lattice.

\subsection{Comparison of the various designs \label{sec:CompVarDes}}

We first compare designs (a) to (d), since design (e) was optimized
with a different cost function. There is a strong degree of
correlation between the hole radii for all four instruments. The dip
in the radius progression between holes 10 and 14 is a common feature,
as is the tendency for the holes to become progressively larger when
their distance with the mouthpiece increases. There is, however, a
significant difference in radius regularity between the
instruments. The designs (a) and (c) without the tuning enlargement
are more regular than (b) and (d), especially (b). Similar
observations can be made regarding the lengths of the chimneys, but
one notices that the constraint on these lengths is effective for
several of the holes for (a) and (c), but not so for (b) and (d). The
situation concerning the positions of tone holes is slightly
different.

\begin{figure}[tbp]
\begin{center}
\leavevmode
\includegraphics[width=11cm]{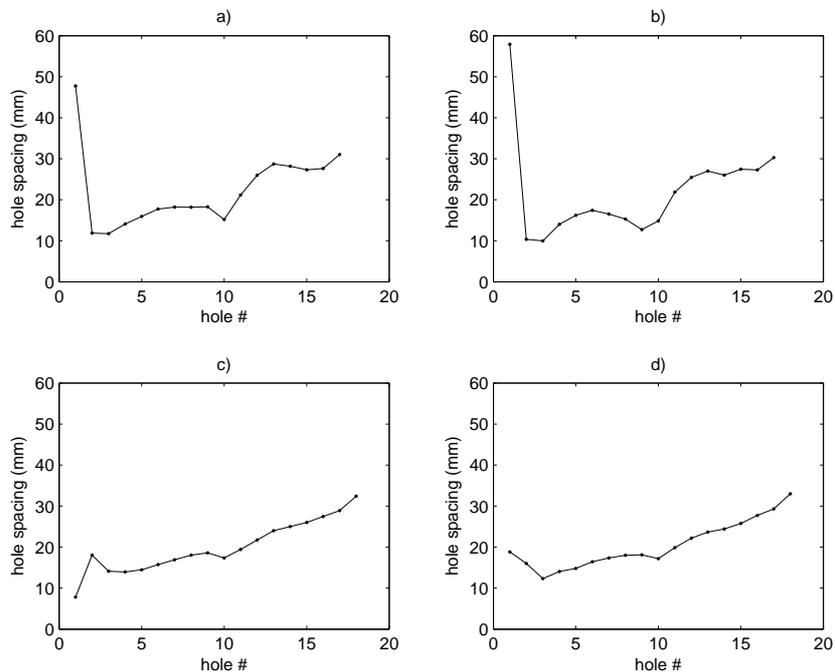}
\end{center}
\caption{Distances between adjacent tone holes for
instruments (a)--(d). }
\label{fig:hole_dist}
\end{figure}

Fig.~\ref{fig:hole_dist} shows the distances between adjacent tone
holes of the four designs.  Here, (a) and (b) are similar, as are (c)
and (d): the designs with a separate register hole are more regular
than the ones with a dual register hole, but the introduction of a
tuning enlargement does not seem to have any adverse effect on the
regularity of hole positions. Design (b) and (c) are roughly
comparable in terms of intonation, but the latter has a much smoother
tone hole pattern.  The conclusion is that, if the bore enlargement
improves the tuning of an otherwise cylindrical instrument, the price
to pay is a less regular tone hole pattern.

The long register hole of (b) and (e) are significantly different from
those of a regular clarinet. At low sound levels, with a linear
behavior, the main effect of a register hole is reactive (the ratio of
the boundary layer thickness to the radius remains small). Therefore
the main parameter is the shunt acoustic mass, proportional to the
ratio length/cross-section area. In order to have a small
perturbation, it is necessary to have a large acoustic mass, therefore
either a long chimney or a small radius. For practical reasons, a very
small radius is not suitable, so that a long chimney is required. This
seems to be what is happening here, since the optimization leads to
the minimum allowed value of the radius (1 mm). The obtained length of
the height of the register hole is unusually large; when the hole is
open, a problem is the insufficient reduction of the heights of the
first impedance peaks .

Concerning design (e), we note that the results provide more
regularity in the geometry of its holes than the others.  In
optimization, it is well-known that under-determined problems may
easily lead to irregular solutions.  Indeed, in this case, it seems
that putting more constraints on the optimization (by including the
position of the resonance of the upper notes of the second register)
leads to more satisfactory results.

To summarize, the best method to obtain intonation, as well as
regularity, seems to be the introduction of a separate register hole.
If a separate register hole is used, an enlargement is not necessary
in order to achieve an instrument that is in tune within 8 cents for a
29 notes range. Design (c) seems to be a good compromise, with most
resonances falling very close to the target (differences of less than
5 cents).  In addition, this design is not very different from a
standard clarinet, even if it is significantly more regular and
requires no cross fingering. But this optimization does not correspond
to a fundamental limit: if, for instance, more deviations from a
cylindrical bore were permitted, it would probably become possible to
adjust resonance frequencies even more accurately.

\subsection{Convergence properties of the optimization \label{sec:convstudy}}
\begin{figure}[tbp]
\begin{center}
\leavevmode
\includegraphics[width=11cm]{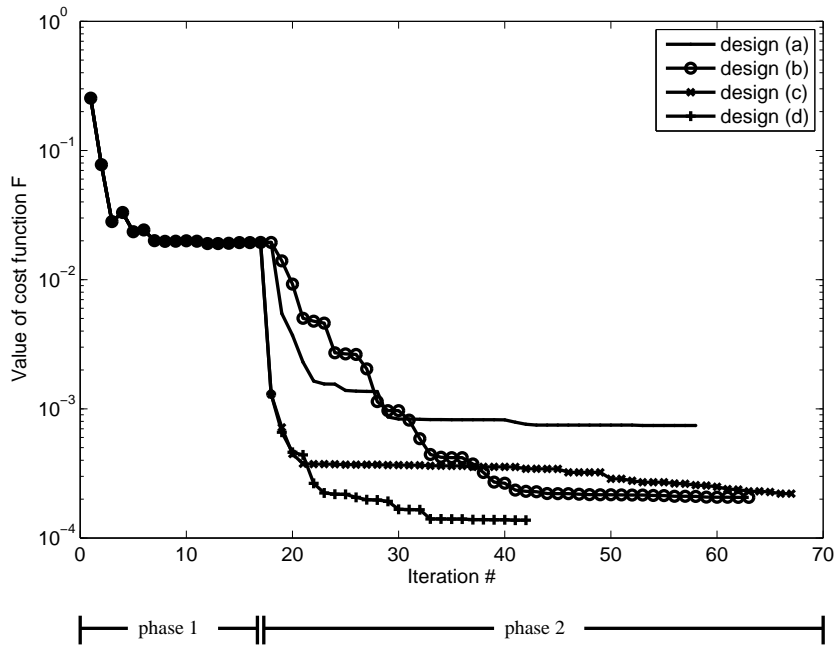}
\end{center}
\caption{Evolution of the target function $F_2$ associated with the
resonance frequencies of the notes or two registers, as a function of
the number or iterations. The first point corresponds to a
non-optimized design used as a seed for the iteration. During phase 1,
the cost function used was $F_1$, which takes into account only the
resonance frequencies of the first register. In phase 2, the cost
function was changed to $F_2$. The whole process provides a reduction
of the target function by a factor ranging from 100 to 1000.}
\label{fig:convcurve}
\end{figure}

It is interesting to study the evolution of the target
function $F$ for the two successive optimization phases: optimization
with a target function $F_1$ that takes into account only the 19 notes
of the first register and optimization with a function $F_2$ that
includes the notes of the two registers. Fig.~\ref{fig:convcurve} shows
how $F_2$ evolves as a function of the number of iterations, for the
designs (a)--(d). The initial point of the curves corresponds to the
crude initial design of the instrument, which is accurate within 12
cents (RMS) throughout the first register, but totally out of tune in
the second register. A rapid decrease of function $F_2$ is observed,
which saturates to a plateau after a few steps. At the end of this
process, a RMS error of 6.5 cents is obtained for the first register,
and of 43 cents for the second. When the optimization function is changed
from $F_1$ to $F_2$, a new rapid decrease of the target function takes
place; this is not surprising since the function shown in the figure
now corresponds exactly to the target function used in the
optimization. Finally, the values of the errors obtained in
Figs.~\ref{fig:CylNoreg}--\ref{fig:EnlargReg} are obtained. Little
success was achieved omitting phase 1. In the rare cases the process
converges at all, convergence is slow with an erratic evolution of the
target function.

In order to escape possible local minima, we have also
used stochastic optimization procedures by adding random perturbations
to the previous optimized designs, and using them to run the
deterministic optimization procedure. We did not obtain significant
improvement of the results in this way. Further studies are probably
needed to better understand the optimization properties of the target
functions associated with tone hole patterns.

\section{Experimental prototype}\label{sec:prototype}
We chose to build configuration (e) obtained in the previous section,
since it offers more regularity in its design.  In order to keep the
fabrication process as simple as possible, stock polyurethane tubes
were used, and no attempt was made to build keys. These tubes come in
a limited set of dimensions, of which the one that is closest to a
real clarinet has a nominal inner diameter of 14.25 mm. This
corresponds to the diameter chosen in the optimization of design
(e). The tube did not show a perfectly circular cross section, but had
a diameter varying between 14.10 mm and 14.45 mm, a non-negligible
variation. Ref. \cite{dekela:05} shows that the corresponding length
correction is bounded by the following equation (Eq. 31 of that
reference):%
\begin{equation*}
\left\vert \Delta \ell \right\vert <(1-\alpha )\ell ^{\prime }=0.05\ell
^{\prime }
\end{equation*}
where $\alpha =S_{\min }/S_{\max }\simeq 0.95$ and $\ell ^{\prime }$
is the length of the enlargement.  The tube was modeled as a cylinder
with the same cross section area, which corresponds to a diameter of
14.27 mm. The chimney lengths were adjusted by creating a flat
external surface at the position of each tone hole, which is drilled
perpendicularly to the main axis of the tube. The edges of the holes
are kept sharp, a feature that may potentially introduce nonlinear
flow effects at high playing levels. Fig. \ref{fig:plasticprototype}
shows the prototype. In Appendix, a workshop drawing
of the instrument is presented.

\begin{figure}[htb]
\begin{center}
\leavevmode
\setlength{\unitlength}{1mm} %
\includegraphics[width=8cm]{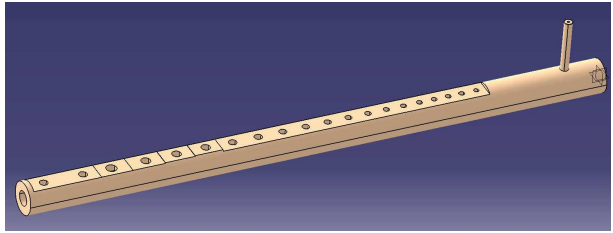}
\end{center}
\caption{The prototype.}
\label{fig:plasticprototype}
\end{figure}

\begin{figure}[tbp]
\begin{center}
\leavevmode
\includegraphics[width=6cm]{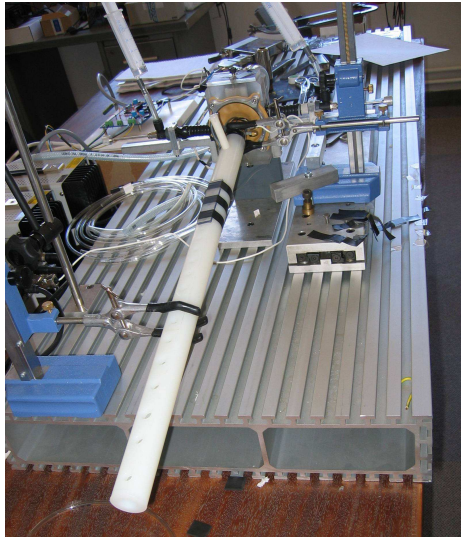}
\end{center}
\caption{The prototype instrument attached to the artificial
mouth.}
\label{fig:prototype}
\end{figure}

The prototype was blown with an artificial mouth (see
Fig. \ref{fig:prototype}), with a standard mouthpiece and a
\textquotedblleft Plasticover\textquotedblright\ reed.  A preliminary
calibration of this device was necessary to measure the equivalent
volume of the mouthpiece/reed ensemble. This volume is used to
calculate the length of upstream cylindrical tube that was removed
from the results of optimization in order to build the prototype. The
measurement was made experimentally by fitting the mouthpiece to a
cylindrical piece of tubing terminated by an orifice in a large
baffle, and deriving a length correction from the measured oscillation
frequency. To check consistency, the experiment was repeated with
different tube lengths and blowing pressures.
\begin{figure}[tbp]
\begin{center}
\leavevmode
\includegraphics[width=10cm]{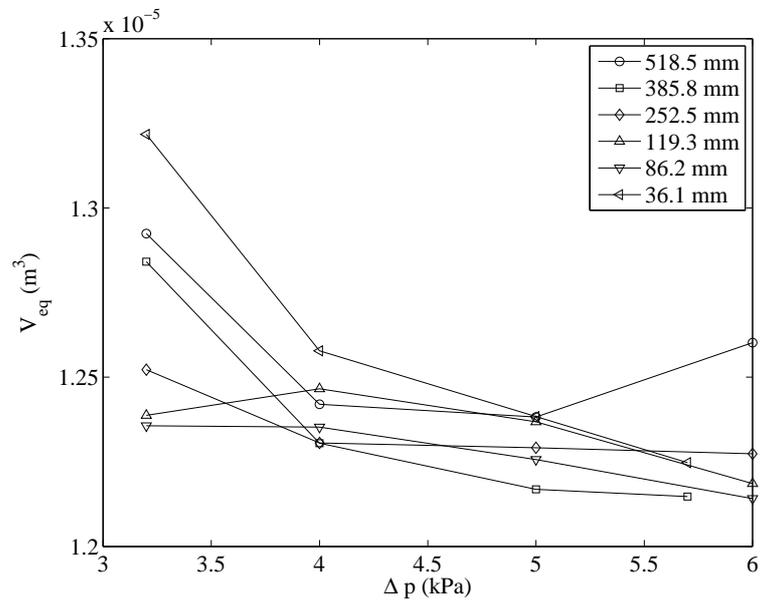}
\end{center}
\caption{Equivalent volume computed from played frequency for different tube
lengths and blowing pressures; 1 kPa corresponds to a water column of 10 cm.}
\label{fig:equivol}
\end{figure}
The blowing pressure was varied from the oscillation threshold to the
saturation limit at which the reed closes against the mouthpiece and
blocks the oscillation. Fig.~\ref{fig:equivol} shows the results,
ranging from 12.2 to 13.2 cm$^3$, to be compared to the geometrical
volume of the mouthpiece (11.4 $\pm$ 0.3 cm$^3$). Since the variation
of the equivalent volume are larger with low blowing pressures, in
order to minimize nonlinear effects, a working pressure of 4 kPa
(about 40 cm of water) was chosen, with $V_{eq}=12.5$ cm$^3$. This
volume corresponds to a tube length correction of about 73 mm.

\begin{figure}[tbp]
\begin{center}
\leavevmode
\includegraphics[width=10cm]{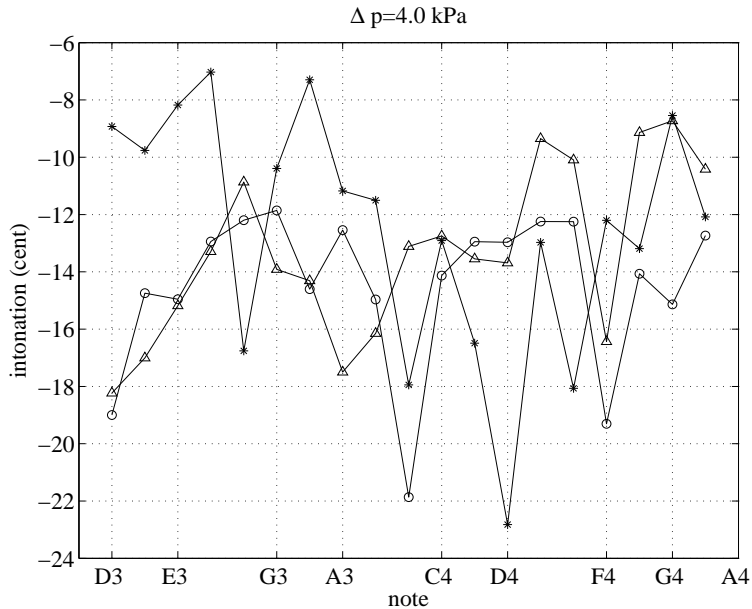}
\end{center}
\caption{The experimental clarinet: measured intonation errors with a
blowing pressure of 4.0 kPa in the artificial mouth
(three realizations).}
\label{fig:intonation40}
\end{figure}

\begin{figure}[tbp]
\begin{center}
\leavevmode
\includegraphics[width=10cm]{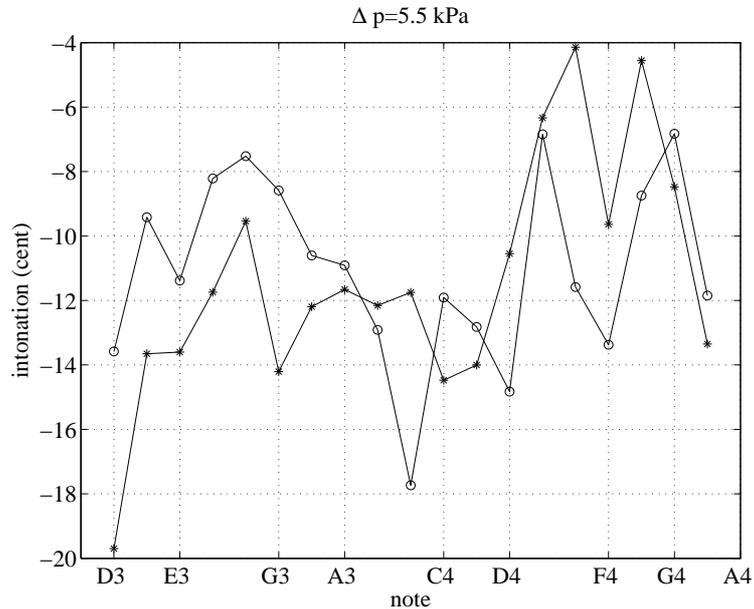}
\end{center}
\caption{The experimental clarinet: measured intonation errors with a blowing pressure of 5.5 kPa
(two realizations).}
\label{fig:intonation55}
\end{figure}
The prototype was then studied. The tone holes were successively
closed with tape on which rigid plastic pads were placed, in order to
replace the pads and keys.  Figure \ref{fig:intonation40} shows the
results obtained with a blowing pressure of 4.0 kPa, and three series
of measurements.  From one series to the next, the instrument is
removed from the artificial mouth. Care was taken to try and
obtain as much reproducibility as possible, but it is clear that this
reproducibility was not perfect, which probably explains the
dispersion of the results.  For the first series, the average of sound
frequencies is 12 cents too low, with a mean square deviation of 4;
for the second, the average is 14 cents too low, with a mean square
deviation of 2.7; for the third, the average is 13 cents too low, with
a mean square deviation of 2.9. Fig. \ref{fig:intonation55} shows
similar results with a blowing pressure of 5.5 kPa.  The first series
of measurements give an average 11 cents too low, with a mean square
deviation of 3.6, the second, an average also 11 cents too low with a
mean square deviation of 2.8. As can be seen, there is a significant
dispersion of the results. The reason for this dispersion is that,
from one run of the experiment to the next, adjustments of the
experimental parameters turned out to be necessary.  The general
offset of the pitch, approximately 10 cents flat, is easy to correct
by adjusting the length of the instrument, as routinely done by
instrumentalists.  This offset being ignored, the remaining errors are
less than 5 cents, which is better than what is usually obtained with
real clarinets.
\begin{figure}[h]
\centering \includegraphics[width=10cm]{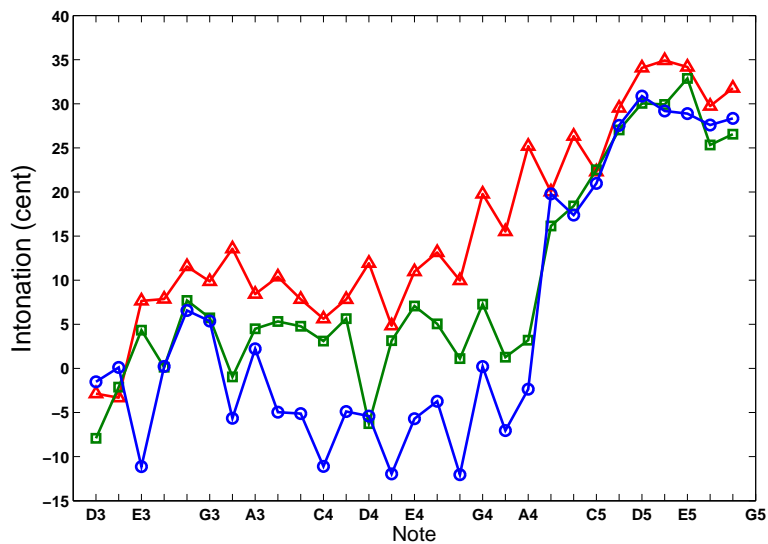}
\caption{Measured intonation errors (in cents) when the experimental
clarinet was played by a musician. Each note was played three times,
at different intensities. Different symbol sizes are used for the
three intensities: $\bigtriangleup$ Piano; $\square$ Mezzoforte; O
Forte - color online}
\label{figalexis}
\end{figure}

For the second register, stable sounds were difficult to obtain with
the artificial mouth. A musician was therefore asked to play the
prototype. Attacking each note, she played the notes of the two
registers successively, but also observed that the second register was
less stable than with an usual clarinet. For each register, she played
the higher notes by closing the holes with the fingers, and the lower
notes by closing the 8 upper holes with modeling clay. In a
preliminary experiment, the general intonation was too low (roughly 30
cents, with a rather unsatisfactory balance between the two
registers); this is not so surprising since the mouthpiece used by the
instrumentalist was not the same as that of the artificial mouth. The
experiment was then slightly modified by reducing the volume of the
mouthpiece by an equivalent length of 1 mm, using modelling clay; the
results are shown on Fig. \ref{figalexis}. Intonation is slightly
higher than that obtained with the artificial mouth, but the agreement
remains rather satisfactory, as well as reproducibility. The pressure
in the mouth was measured to be between 4 and 5 kPa, as for the
artificial mouth. Between the two registers, a discontinuity of 20
cents can be observed. This can be due to the playing technique of the
instrumentalist. It seems likely that between the two registers, she
probably changed the excitation parameters, such as the reed opening
and the mouth pressure. Moreover, no listening reference was given
before she played the note; the player just optimized easy
playing. Usually, measurements of the intonation of a clarinet is made
in less severe conditions, where the musician plays all notes in
succession so that he can keep a reference in mind and automatically
apply pitch corrections. Generally speaking, it turned out that all
notes could be played without any special training, which is rather
satisfactory.

\begin{figure}[tbp]
\begin{center}
\leavevmode
\includegraphics[width=10cm]{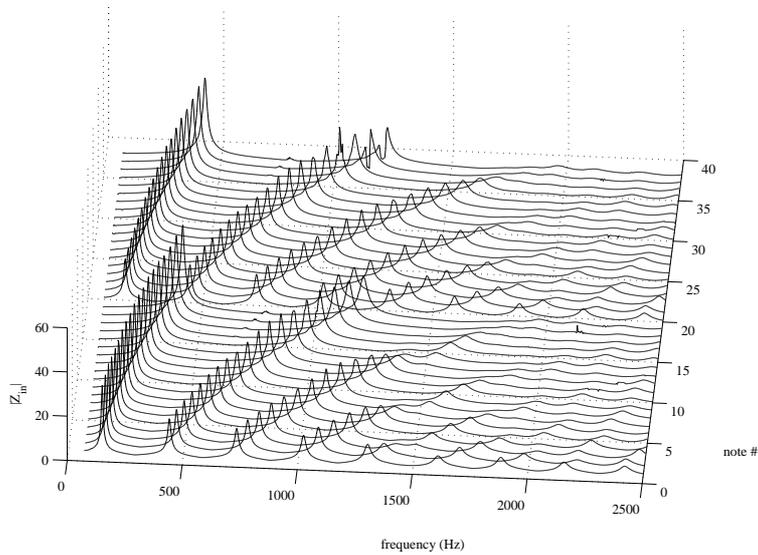}
\end{center}
\caption{Measured input impedance spectrum of the experimental clarinet.
This figure shows good agreement with the computed results of
Fig. \ref{fig:expclar1WF}.}
\label{fig:measimpcascade}
\end{figure}

Comparisons between calculated
(Fig.~\ref{fig:expclar1WF}) and measured
(Fig.~\ref{fig:measimpcascade}) impedances on the prototype for the 36
semi-tones show an average deviation on the resonance frequencies of 1.97 cents while the standard
deviation is 11.3 cents. It is worth noticing that the last notes of
the second register contributes to this result.
In disagreement with the assumption that tone hole
interaction is small, the spacing between some of the tone holes is
slightly smaller than the bore diameter. This may account for some of
the deviation between the model and the behavior of the prototype.

\section{Conclusion}
\label{sec:discussion}

Computer optimization of the geometry of a clarinet seems to offer
interesting possibilities, even if it should be remembered that the
numerical results do not necessarily correspond to an absolute optimum
for the chosen criterion: they may be only local optima. The
regularity of the obtained geometries seems to indicate that, indeed,
the design of real instruments is more the result of a complicated
history than that of pure logics.  For the moment, our study remains
limited in terms of the number of acoustical properties taken into
account in the optimization function, since only the positions of the
acoustical resonances have been included. It would be interesting to
also include the corresponding value of the impedance peaks, which
might lead to significantly different optimization results.  Even if
the results seem to be satisfactory in terms of the peak values of the
acoustic impedance, the relative heights of the peaks is important;
for instance, the differential reduction of the heights of the first
and second resonance determines the stability of emission for the
second register.

Generally speaking, there should be no special difficulty in including
more components in the optimization function, but our purpose in the
present work was to explore the new possibilities offered by
optimization within the simplest possible scheme; experience will show
in what direction the optimization process should be
improved. Moreover, it remains very likely that even a very elaborate
mathematical optimization model will probably never capture all the
real musical possibilities of instruments. At some point, it will be
indispensable to build playable instruments with keys and collect the
evaluation of performing clarinetists; mathematical optimization can
nevertheless be very useful as a preselection tool between the
enormous number of geometrical possibilities, even if its use should
be followed by a final adjustment with real musical testing by
performers. We hope to be able to continue our program in this
direction.

\section{Acknowledgments}

We wish to thank Alain Busso, Didier Ferrand for their work on the
experimental instrument and the artificial mouth, and Camille-Eva
Rakovec for playing the prototype. This work has been supported by
the French ANR research project CONSONNES, and more recently by the
ANR project CAGIMA.

\section{Appendix}
\begin{figure}[h!]
\begin{center}
\leavevmode
\includegraphics[width=8cm,angle=90]{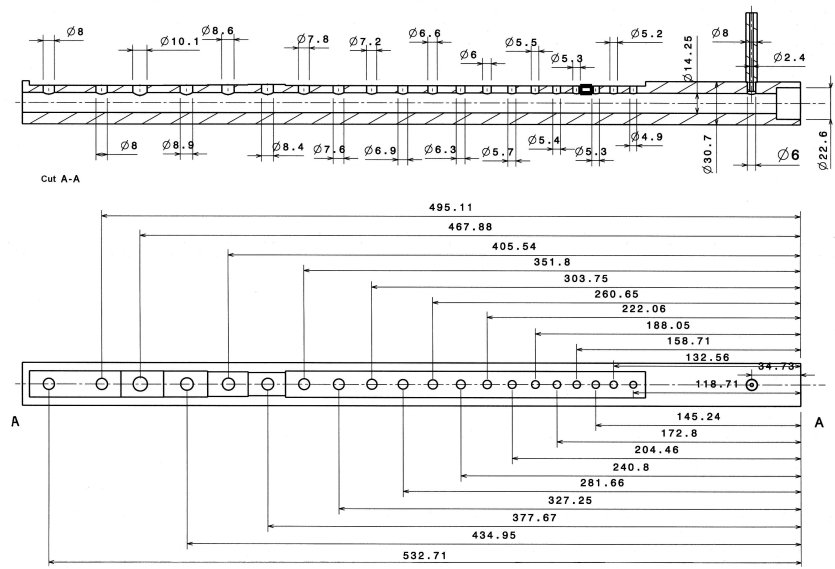}
\end{center}
\caption{Drawing of the prototype instrument.}
\label{fig:drawing}
\end{figure}

\end{document}